# Influence of oxygen in architecting large scale nonpolar GaN nanowires

Avinash Patsha,[1,*] S. Amirthapandian,[2] Ramanathaswamy Pandian[1] and S. Dhara[1,*]

[1]Surface and Nanoscience Division, Indira Gandhi Center for Atomic Research, Kalpakkam-603102, India

[2]Materials Physics Division, Indira Gandhi Center for Atomic Research, Kalpakkam-603102, India.

**Corresponding Authors:** avinash.phy@gmail.com; dhara@igcar.gov.in

*Abstract*

Manipulation of surface architecture of semiconducting nanowires with a control in surface polarity is one of the important objectives for nanowire based electronic and optoelectronic devices for commercialization. We report the growth of exceptionally high structural and optical quality nonpolar GaN nanowires with controlled and uniform surface morphology and size distribution, for large scale production. The role of O contamination (~1-$10^5$ ppm) in the surface architecture of these nanowires is investigated with the possible mechanism involved. Nonpolar GaN nanowires grown in O rich condition show the inhomogeneous surface morphologies and sizes (50 - 150 nm) while nanowires are having precise sizes of 40($\pm$5) nm and uniform surface morphology, for the samples grown in O reduced condition. Relative O contents are estimated using electron energy loss spectroscopy studies. Size-selective growth of uniform nanowires is also demonstrated, in the O reduced condition, using different catalyst sizes. Photoluminescence studies along with the observation of single-mode waveguide formation, as far field bright violet multiple emission spots, reveal the high optical quality of the nonpolar GaN nanowires grown in the O reduced condition.

# Introduction

Group III nitride based GaN nanowires have emerged as potential building blocks for nanoscale electronic and optoelectronic devices such as LED, lasers, high electron mobility transistor (HEMT), logic gates, photodetectors, solar cells and gas sensors.[1-10] However, in large scale production control over the performance of such nanowire devices involves the issues ranging from growth to processing of their building blocks, i.e. nanowires. Efforts have been made to understand the variations in growth and various electrical and optical transport properties of nanowires with respective to their size, dopants, defects and crystallographic orientations.[11-14] Controlling diameter and surface morphology of GaN nanowires play a crucial role in defining electrical, optical and electromechanical properties, for using such building blocks for LED, HEMT, chemical sensor and nanoelctromechanical sensor (NEMS) applications.[15,16].

In the conventional polar GaN (*c*-axis oriented) materials, polarization induced electric fields parallel to the growth direction causes the reduction in external quantum efficiency of white light emitting diodes.[17] Whereas the nonpolar GaN (oriented along $[10\bar{1}0]$ and $[11\bar{2}0]$ ) materials have attained intense research interest due to the absence of polarization induced electric fields.[18,19] The effects of dopants and impurities either on the defect formation or on the evolution of surface morphology is very important issue in the device performance and have been well studied in case of thin films of GaN samples.[20-22] The low energy nonpolar planes of GaN are being affected by defects and impurities leading to formation of pinholes and nanopipes in thin films.[23-25] While a material is scaled down to nano-dimensions, particularly in bottom-up approach, it is important to preserve the surface architecture of nanostructures in the entire sample. Apart from variations in properties due to presence of impurities, excessive unintentional incorporation of them into these nanostructures may also lead to the devastation of architecture of the nanostructures.

The most studied ubiquitous inevitable impurity in group III nitrides is the O which impoverishes the nitride phase and alters the electrical, optical and electromechanical properties. The degree of O contamination depends on the growth technique.[26,27] The common source of this contamination in chemical vapor deposition (CVD) based technique is known to be reactant $NH_3$ and carrier gases, where as in molecular beam epitaxy (MBE) it is remnant water vapor.[28] Apart from these sources, a background level of O can be liberated from the substrates and growth chamber like sapphire and quartz tubes. The consequences of O contaminations are more in case of atmospheric pressure CVD technique as the base vacuum of the chamber and purging precursors are inadequate to reduce the O concentration to background level. In spite of these problems, the extent over which the effect of unintentional incorporation of O in GaN nanowires is unexplored. Although carbon impurity influences the optical properties by its electronic contribtion,[29] but unlike oxygen it does not directly participate in the VLS reaction process with Ga. However, there is hardly any report showing morphological changes with carbon contaminations. The reproducibility and homogeneity of required size and morphology of GaN nanowires in a simple CVD technique is a critical issue for mass production of ensembled nanowire devices.

In the present study we have investigated the effect of unintentional incorporation of O on the nonpolar GaN nanowire growth, size and their morphology. We have shown that the excessive incorporation of oxygen into the GaN during the nanowires synthesis could lead to the uncontrollable growth rate and inhomogeneous surface morphologies. By reducing the O contamination to a critical level, we were able to synthesize and reproduce size selective growth of nanowires with uniform size and shape. Along with the perfection in architectural features, we have also demonstrated improvement in the optical properties of the nanowires in the formation of waveguide showing far field emission.

**Experimental Methodology and Characterization Techniques**

GaN nanowires were synthesized in atmospheric pressure chemical vapor deposition technique using vapor-liquid-solid (VLS) process. Au islands were deposited on Si(100) substrates in a thermal evaporation technique (12A4D, HINDHIVAC, India). These substrates were separately annealed for making the Au nanoparticles at a temperature of 900 oC for 15 min in the inert atmosphere. We have used Ga metal (99.999%, Alfa Aesar) as a Ga source, $NH_3$ (3N and 5N pure) as reactant gas, Ar (commercial) and mixture of ultra high pure (UHP) $Ar+H_2$ (5N) as carrier gases. Si substrate with Au nanoparticles is kept upstream to a Ga droplet in a high pure alumina boat (99.95%) which was placed into a quartz tube. The temperature of the quartz tube was slowly raised to a growth temperature of 900 $^{o}$C with 15 $^{o}$C/min ramp rate. Nanowires were grown for 60 min growth time by purging 10 sccm of $NH_3$ reactant gas and 20 sccm of Ar carrier gas. Incorporation of O has been carried out at different oxygen partial pressures varied between $1.2 \times 10^{-4}$ Torr to $1.3 \times 10^{2}$ Torr by varying the base pressure of the growth chamber (quartz tube) before the growth time and gas purity of $NH_3$ and Ar. For O rich condition, nanowires were grown at an oxygen partial pressure of $1.3 \times 10^{2}$ Torr (~$2 \times 10^{5}$ ppm $O_2$; sample R1) by purging high pure $NH_3$ (3N) and commercial Ar gases without evacuating the chamber. For O reduced condition, two sets of nanowire samples were grown at oxygen partial pressures of $1.2 \times 10^{-1}$ Torr (~150 ppm $O_2$; sample R2) and $1.2 \times 10^{-4}$ Torr (< 1 ppm $O_2$; sample R3) by purging ultra high pure $NH_3$ (5N) and UHP $Ar+H_2$ (5N) respectively. All other growth parameters like average size of the catalyst Au nanoparticles, temperature, flow rate and growth time were kept constant during the growth of three sets of samples mentioned above. Another set of sample at an oxygen partial pressure of $1.2 \times 10^{-4}$ Torr (< 1 ppm $O_2$; sample R4) is also grown using catalytic Au nanoparticles of different average size than that used in previous case for realizing size control in the growth process.

Morphological features of the as grown samples were examined by field emission scanning electron microscope (FESEM, SUPRA 55 Zeiss). For structural studies, high resolution transmission electron microscopy (HRTEM, LIBRA 200FE Zeiss) observations were performed on nanowires which were dispersed in isopropyl alcohol and transferred to TEM Cu grids. Electron energy loss spectroscopy (EELS) studies were carried out for identifying N and O concentrations in a single nanowire using in-column $2^{nd}$ order corrected omega energy filter type spectrometer with energy resolution of 0.7 eV. For optical quality investigations, nanowires were excited with UV laser of wavelength 325 nm and collected the photoluminescence (PL) spectra at 80 K using a spectrometer (inVia Reinshaw microscope) with 2400 gr/mm grating and thermoelectric cooled CCD detector.

## Morphological and Structural Analyses

Uniform shape and well separated Au nanoparticles show the size distribution of 30(±5) nm (supporting information Fig. S1). Nanowires grown in O rich condition (R1) show quite rough and non-uniform surface morphology with large size distribution (Figs. 1a, 1b) and large growth rate (17 μm/h). In oxygen reduced condition, nanowires of sample R2 show smooth but non-uniform size distribution (Fig. 1c) while that of sample R3 shows smooth and homogeneous surface morphology (Fig. 1d) and uniform size distribution all over the sample with a reduced growth rate (3 μm/h). The particle at the tip shows that the nanowires are grown in VLS process (Fig. 1d). Although all the three sets of samples (R1, R2 and R3) are grown with similar size catalyst particles, large variation in the diameter of the nanowires of the three samples is observed. Nanowires of sample R1 are having large diameter and non-uniform sizes (50 nm - 150 nm) and non-uniform surface morphologies (Figs. 1a, 1b), whereas the diameter of the nanowires of sample R2 (Fig. 1c) is found to be significantly smaller than that of R1 but with non-uniform sizes (40 nm – 90 nm). We have also studied morphology of the nanowires grown at different oxygen partial pressures between $10^{-4}$ and 90 Torr (~1-100 ppm $O_2$) showing non-uniformity similar to R1 and R2 (supporting information Fig. S2). Surface morphology along the wires, however was improved compared to that of the sample R1. In the sample R3, the diameter of the nanowires exactly follows the size of the catalyst particle at the tip and having uniform shape and size distribution ~ 40($\pm$5) nm with homogeneous surface morphology along the wire, in the entire sample (Fig. 1d). The reproducibility of the uniform shape, size distribution and surface morphology of the sample R3 is improved in the O reduced condition than that found for the samples (R1 and R2) grown in O rich condition. Au nanoparticles of size distribution ~ 40($\pm$5) nm (supporting information Fig. S3) are used for demonstrating size selective growth of nanowires, sample R4, with uniform shape and surface morphology with bigger sizes of 55($\pm$5) nm (Fig. 2). Deposition conditions same as sample R3 is used for this purpose. The uniform size and morphology of the nanowires might be a result of the controlled and thermodynamically stable growth of the crystallites with specific crystalline orientation by the incorporation of Ga and N species in the reduced impurity condition.

In the HRTEM morphological analyses, a typical nanowire of the sample R1 (grown in O rich condition) shows an irregular surface morphology (Fig. 3a). Fast-Fourier transform (FFT) spots corresponding to HRTEM micrograph of from the nanowire (outset of Fig. 3a) reveals presence of mixed phases of GaN and $Ga_2O_3$. The diffraction spots enclosed by doted circles are indexed to $\{100\}$ and $\{\bar{3}11\}$ planes corresponding to monoclinic $Ga_2O_3$ phase and spots enclosed by doted squares are indexed to $\{0002\}$ planes of wurtzite GaN phase. High resolution image of the nanowire shows non-uniform contrast (Fig. 3b) which may originate from variation in thickness along the nanowire and barely seen for lattice fringes. Lattice fringes with d-spacing 0.297 nm and 0.275 nm corresponding to {100} planes of $Ga_2O_3$ phase and {10$\bar{1}$0} planes of GaN phase, respectively are observed in the magnified view (Fig. 3c) of high resolution image. It is observed that (100) planes are presumably grown over nonpolar planes (10$\bar{1}$0) of GaN. Nanowires of the sample R2 is having uniform rod like shape (Fig. 3d) and the corresponding selected area electron diffraction (SAED) pattern shows perfect single crystalline feature (outset of Fig. 3d). The diffraction spots are indexed to [0001] zone axes of wurtzite GaN. Nanowire is having nearly uniform contrast (Fig. 3e) except on one edge of the wire where a thin amorphous layer is observed which may be due to possible formation of an oxide overlayer on the surface of nanowire. Magnified view of lattice resolved micrograph (Fig. 4f) shows the interplanar spacing of 0.276 nm corresponding to nonpolar {10$\bar{1}$0} planes of wurtzite GaN and growth direction of the nanowire is determined to be along [11$\bar{2}$0]. A typical nanowire of the sample R3 grown in O reduced condition shows a perfect rod like shape with uniform surface morphology (Fig. 4a) and having an Au catalyst particle at the tip. The SAED pattern (Fig. 4b) reveals that the nanowire is single crystalline wurtzite phase of GaN with zone axes [0001]. A sharp interface between GaN nanowire and Au catalyst particle is seen in high resolution electron micrograph (Fig. 4c). Interplanar spacing of 0.276 nm, shown in the magnified view (Fig. 4d), corresponds to the nonpolar {10$\bar{1}$0} planes of wurtzite GaN. The growth direction of the nanowire is found to be along [11$\bar{2}$0] direction, similar to that of the sample R2.

The O incorporation in to the nanowire may occur through two possible ways. It may occur through liquid/solid interface (Au-Ga/GaN) while the nanowire is growing in VLS process and incorporation may occur directly on the active nonpolar surfaces of the nanowire. In the O rich condition, when the nanowire grows along a nonpolar direction [11$\bar{2}$0], there is a large tendency of incorporation of O along with N at liquid/solid interface. Thus, a supersaturation mixture of GaN and Ga$_2$O$_3$ phases may have formed bringing instability in the uniformity of nanowire due to fluctuations among thermodynamically non-equilibrium growth of the different atomic planes by the incorporation of N and O species, simultaneously. As a result, nanowires will end up with non-uniform morphology or shape along the axis of the wire. Although the two nonpolar planes, (11$\bar{2}$0) and (10$\bar{1}$0) have similar surface energies,[30] O atoms have strong tendency to segregate on the (10$\bar{1}$0) planes.[24] In the direct incorporation of O in to (10$\bar{1}$0) surfaces, O replaces the N and forms stable defect complexes, in their stability order 2(O$_N$), V$_{Ga}$–3(O$_N$) and finally it supersaturates to form Ga$_2$O$_3$ at the extreme limits of Ga and O.[23,24] On further incorporation, O mobilizes to a new place and the process of defect formation and oxide over layer will repeat. As a result, nanowire will be thick and having rough surface morphology with oxide over layer. In order to confirm the variations in O and N concentrations in the nanowires, we have carried out the electron energy loss spectroscopy (EELS) studies.

## Influence of Oxygen

EELS spectra of nanowires of the three samples are analyzed by collecting the $k$-edge emission of N and O. The spectra show a drastic variation in the fine structures of N-$k$ and O-$k$ edges (Figs. 5a, 5b). All the $k$-edge spectra are background subtracted based on power law energy dependence ($AE^{-r}$).[31] A clear splitting in the fine structure with poor intensity of N-$k$ edge spectra (Fig. 5a) is observed for the nanowire of sample R1, grown in O rich condition, with irregular surface morphology. The fine structures of N-$k$ edges from the nanowires of the samples R2 and R3, grown in the O reduced condition are broadened and show the increase in N concentration. The O-$k$ edge spectra follows reverse trend for samples R1 to R3. O-$k$ edge spectra obtained from the nanowires of samples R2 and R3 (Fig. 5b) are broad and weak while that of the nanowires of sample R1 (Fig. 5b) shows clear splitting in fine structure and intense peaks. The rise and fall in the intensity of O-$k$ edge and N-$k$ edge, respectively for the nanowires of sample R1 is presumably due to the O substitution for nitrogen in the GaN during the growth of samples in O rich condition.

The variation in the N to O concentration with respect to nanowire surface morphology has been investigated by obtaining the electron energy loss $k$-edge spectra of N and O from the two different positions of the each nanowire of the samples R1 and R3 grown in O rich and O reduced conditions, respectively. Figure 6a shows the power law energy dependent background subtracted $k$-edge spectra of N and O obtained from the position 1 (middle region of the nanowire, thin and rough morphology) and position 2 (tip region of the nanowire, thick and rough morphology) on the nanowire of the sample R1. Figure 6b show background subtracted $k$-edge spectra of N and O obtained from the position 1 (middle region of the nanowire) and position 2 (close to one end of the nanowire) on the nanowire of the sample R3. The ratio of two elemental concentrations can be obtained by calculating the core loss integral intensity of $k$-edge, $I_k(\beta,\Delta)$ for a collection-semi angle $\beta$ over an energy range $\Delta$ beyond threshold and a partial scattering cross-section for that particular threshold edge $\sigma_k(\beta,\Delta)$.[31] Therefore the concentration ratio for N and O is given by,

$$\frac{n_N}{n_O} = \frac{I_{kN}(\beta,\Delta)}{I_{kO}(\beta,\Delta)} \frac{\sigma_{kO}(\beta,\Delta)}{\sigma_{kN}(\beta,\Delta)}$$

The core loss partial scattering cross-sections for $k$-shells of N and O, $\sigma_{kN}(\beta,\Delta)$ and $\sigma_{kO}(\beta,\Delta)$, respectively, are calculated using SIGMAK3 program which is based on hydrogenic model.[31] The core-loss intensity of $k$-shells for N and O, $I_{kN}(\beta,\Delta)$ and $I_{kO}(\beta,\Delta)$, respectively, are calculated by integrating the area under the curve over an energy $\Delta$=50 eV above the corresponding thresholds.

The concentration ratio of N to O ($n_N/n_O$) is measured at the position 1 (Fig. 6a) middle region of the nanowire of the sample R1 is found to be 0.20. Whereas the ratio is reduced to 0.09 when the spectra collected at position 2, tip region of the nanowire with thick and rough surface morphology. The data shows that the inhomogeneous incorporation of O into the crystalline planes of GaN. For the nanowire of the sample R3 (Fig. 6b), $n_N/n_O$ is measured at the position 1 and position 2, middle and one end of the nanowire and are found to be 1.21 and 1.19, respectively, showing dramatic reduction in O concentration and homogeneous incorporation of N into crystalline planes of GaN.

**Optical Properties**

Optical quality of the GaN nanowires of three sets of samples (R1, R2 and R3) grown in O rich and O reduced conditions was investigated using PL spectroscopy. A UV laser of 325 nm wavelength is used to excite the nanowires and collected the emission spectra at 80 K. The luminescence from the nanowires of sample R1 (Fig. 7), grown in O rich condition shows a broad emission around 2.0-2.3 eV which may be attributed to YL band.[33,34] A weak emission around 3.27 eV is also observed. In the GaN, oxygen easily substitutes the N ($O_N$) due to low defect formation energy and forms a shallow donor level.[27] The Ga vacancies ($V_{Ga}$) are deep acceptors and forms complexes with O shallow donors as $V_{Ga}O_N$, which possibly causes the emission of YL band around 2.2 eV.[33] In the O rich conditions particularly with concentration above $10^{20}$ cm$^{-3}$, the O doping increases the $V_{Ga}O_N$ complex formation leading to increase in the intensity of YL band. $A_1$(LO) Raman mode along with its second order mode, originating from electron-phonon coupling induced Fröhlich interaction,[35] are also observed for sample R1. Electron-phonon coupling is predominant in these nanostructures where non-zone centre higher order phonon modes are also observed due to breakdown of symmetry at resonant condition with the excitation of 325 nm. In the reducing O condition the intensity of YL band decreases drastically in the luminescence spectra of nanowires of sample R2 while it vanishes in the collected luminescence spectra from nanowire of sample R3. A second broad peak around 3.27 eV in the spectra of sample R1 (Fig. 7) is attributed to recombination of neutral donor-acceptor pair (DAP; $D^0A^0$), due to transition from a shallow donor state of nitrogen vacancy (VN) to a deep acceptor state of $V_{Ga}$. The DAP peak is red shifted by 10 meV which may be the result of defects present in the nanowire.[35] Emission due to excitons bound to neutral donors ($D^0X_A$) which is commonly peaked at 3.47 eV is completely suppressed.[33] This shows that the optical quality of nanowires of the sample R1, grown in O rich condition, is degraded. The luminescence spectrum of nanowires of the sample R2 (Fig. 7) shows strong emission around 3.0- 3.5 eV with a significant reduction in YL compared to that of sample R1. The spectrum consists of a free exciton (FE) emission around 3.51 with its phonon replica around 3.385 (FE-LO) eV.[33] A peak corresponding to DAP ($D^0A^0$) type transition around 3.27 eV with its phonon replica, $D^0A^0$-2LO at 3.1 eV and a blue luminescence (BL) around 2.85 eV are also observed.[35] Nanowires grown in O reduced condition of the sample R3 emitted a strong band of $D^0X_A$ at 3.47 eV (FWHM ~ 130 meV) with a $D^0A^0$-2LO band around 3.31 eV. Surprisingly we have observed the 2LO coupling in some of the PL spectra while coupling with 1LO mode is absent. The absence of 1LO coupling is not very clear and it may be due to the breakdown of symmetry in the resonant condition. The absence of YL band and strong emission of DX band in the O reduced condition are the implication of high optical quality GaN phase.

We have also used these GaN nanowires, grown in O reduced condition, as waveguide for observing far field excitation-emission process. Figures 8a and 8b show multiple violet emission spots from R3 and R4 samples, respectively at different places on the nanowires grown area recorded at room temperature (300 K) as well as 80 K. Very often we observe bright violet emission spots corresponding to the $D^0X_A$ band around 3.47 eV (357 nm with FWHM 27 nm) of PL emission for R3 sample (Fig. 8). GaN is reported to have very high refractive index of 2.7 at 357 nm wavelength.[37] These emissions are coming out of the ends of the nanowires through the available open space while other ends are at the excitation with laser light. In this regard we have also calculated the condition for a nanowire to function as a single-mode optical waveguide,[38] where $1 \approx \pi d\sqrt{n_1^2 - n_2^2}/\lambda < 2.4$, with 1 being the practical lower limit. For diameter, $d$ of our nanowires ~ 40($\pm$5) nm (sample R3) and 55($\pm$5) nm (sample R4) at wavelength, $\lambda$ =357 nm; $n_1$ ~ 2.7 and $n_2$ = 1 are the refractive indices of the nanowire around 357 nm and surrounding medium of air, respectively, we can get the minimum value close to 1. Observation of only few emission spots (Fig. 8) may be due to the fact that opening of the other end of the nanowire is not available for most of them as they are grown on substrate.

## Summary

Size selective nonpolar wurtzite GaN nanowires with uniform surface morphology are grown by reducing the O concentration to background level. Nanowires, grown in O rich condition, show inhomogeneous surface morphologies and non-uniform size distribution with the presence of both GaN and $Ga_2O_3$ phases. Electron energy loss spectroscopic measurements confirm large variations in the O concentrations along the nanowire grown in O rich condition. High resolution transmission electron microscopic analysis reveals that the nonpolar nanowires are grown along [11$\bar{2}$0] with side facets (10$\bar{1}$0). Far field bright multiple violet emission spots form nanowires, grown in the O reduced condition, show formation of single-mode waveguide in these samples. Thus by controlling the O concentration, large scale growth of high structural and optical quality nonpolar GaN nanowires is reported in a simple yet versatile chemical vapor deposition technique. The results pave the way for commercialization of ensembled GaN nanowire devices for optoelectronic applications.

## Acknowledgment

One of us (AP) acknowledges Department of Atomic Energy for the financial aid. We thank A. K. Sivadasan, MSD, IGCAR for his help in experiments and Jayasurya Basu, Materials Synthesis & Structural Characterisation Division, IGCAR for his useful discussions. We also thank A. K. Tyagi and C. S. Sundar of Materials Science Group, IGCAR for their encouragement and support.

**Electronic Supplementary Material**: Supplementary material (FESEM image of Au catalyst particles for size-selective growth of nanowire) is available in the online version of this article

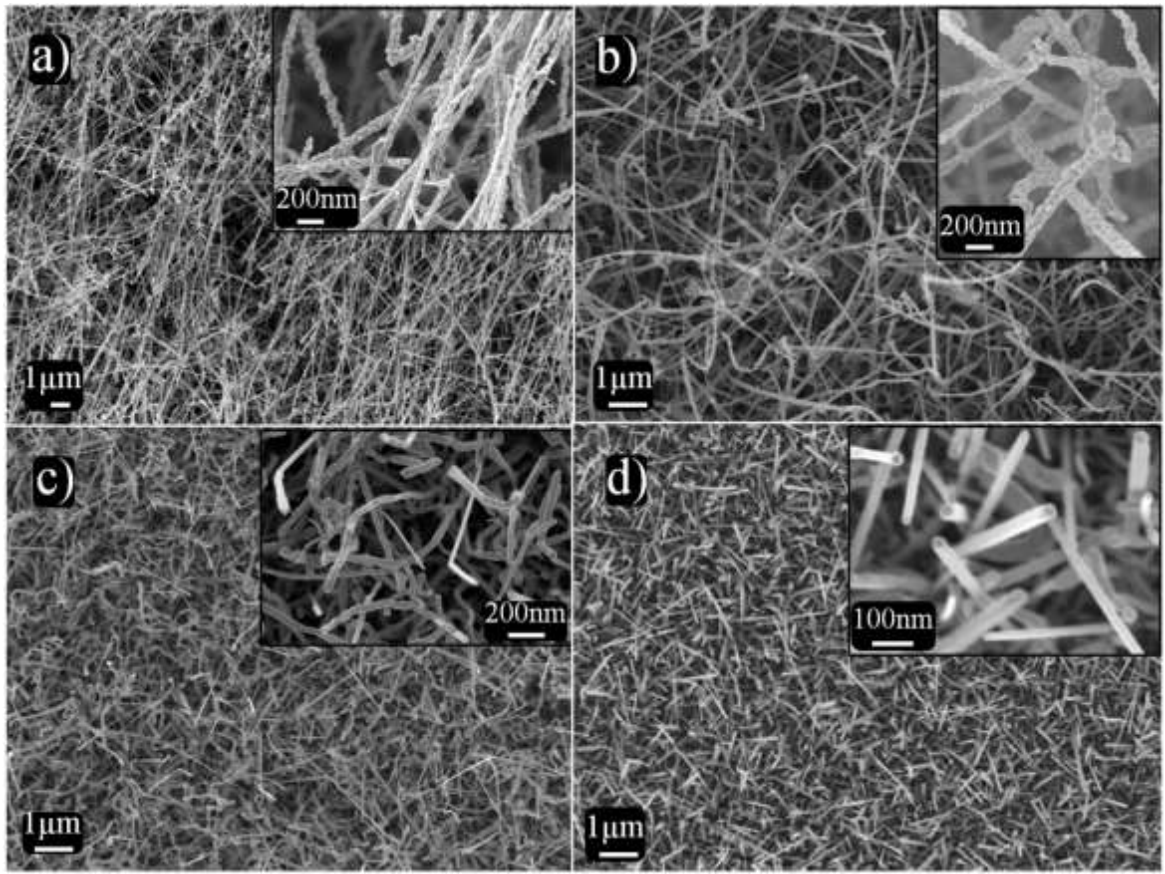

**Fig. 1**. Typical FESEM images of GaN nanowires. a), b) Nanowires grown in oxygen rich condition (~$10^5$ ppm $O_2$; sample R1) showing (insets) quite rough and nonuniform surface morphology in the respective high resolution images. c) Nanowires grown in oxygen reduced condition (~150 ppm $O_2$; sample R2) shows smooth but nonuniform morphology in the respective high resolution image. d) Nanowires grown in oxygen reduced condition (<1 ppm $O_2$; sample R3), having the uniform diameter and surface morphology along the nanowires shown in the high resolution image (inset). Inset, nanoparticles at the tip of the nanowires are clearly seen.

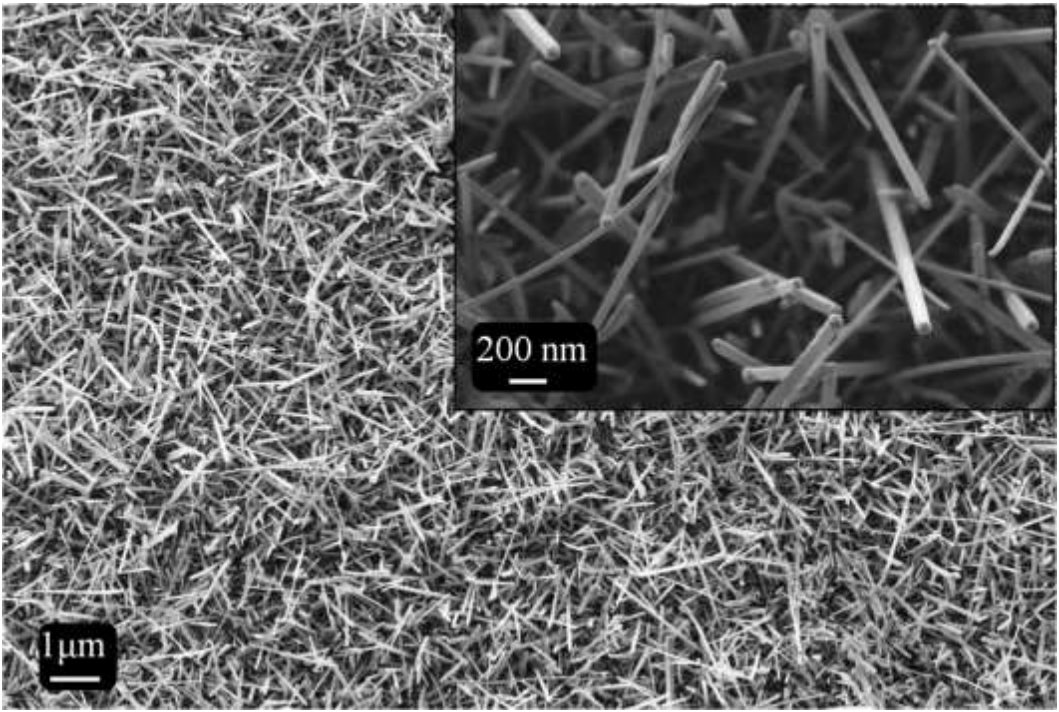

**Fig. 2**. Typical FESEM images of GaN nanowires grown in oxygen reduced condition (<1 ppm O$_2$; sample R4). Inset shows high resolution image of these nanowires with uniform shape, surface morphology and size distribution of 55($\pm$5) nm.

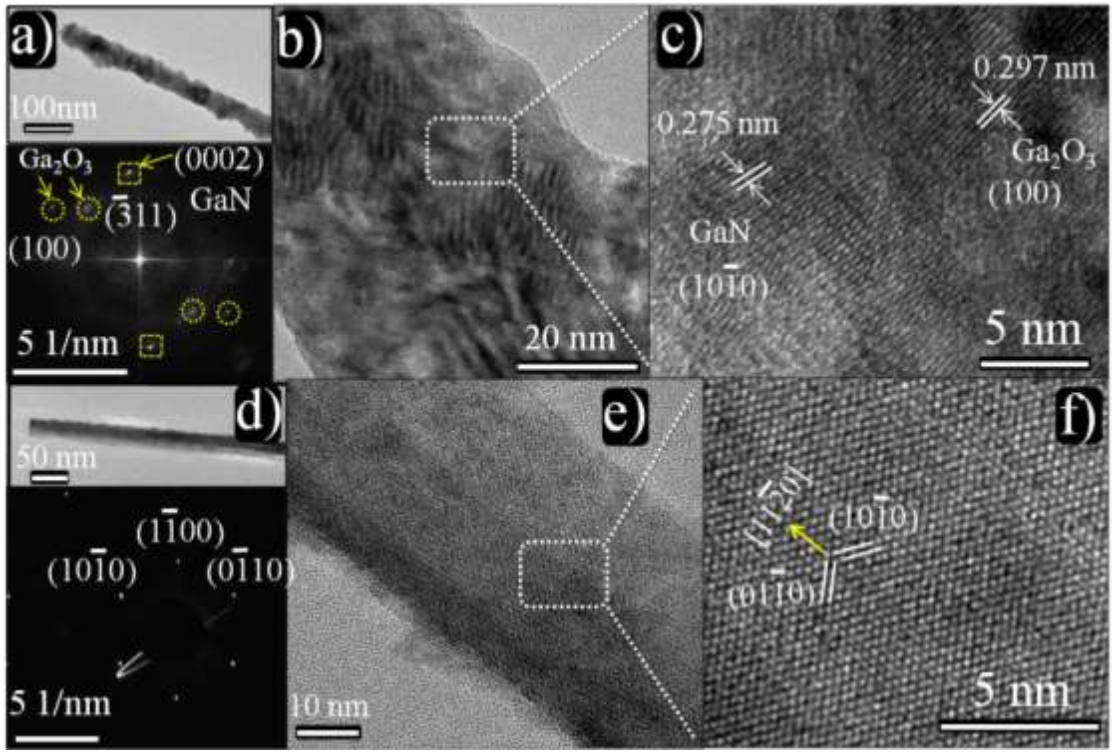

**Fig. 3**. a) A typical nanowire grown in O rich condition (sample R1) shows an irregular surface morphology. Fast-Fourier transform (FFT) diffraction spots corresponding to high resolution TEM micrograph (outset) reveal the presence of phases of GaN and $Ga_2O_3$. b) and c) High resolution TEM images of nanowires. d) A typical nanowire grown in O reduced condition (sample R2) with selected area diffraction pattern (outset), indexed to wurtzite GaN of zone axes along [0001]. e) and f) High resolution micrographs of the nanowire (Fig. 4d) having nonpolar planes $(10\bar{1}0)$ and the growth direction along $[11\bar{2}0]$.

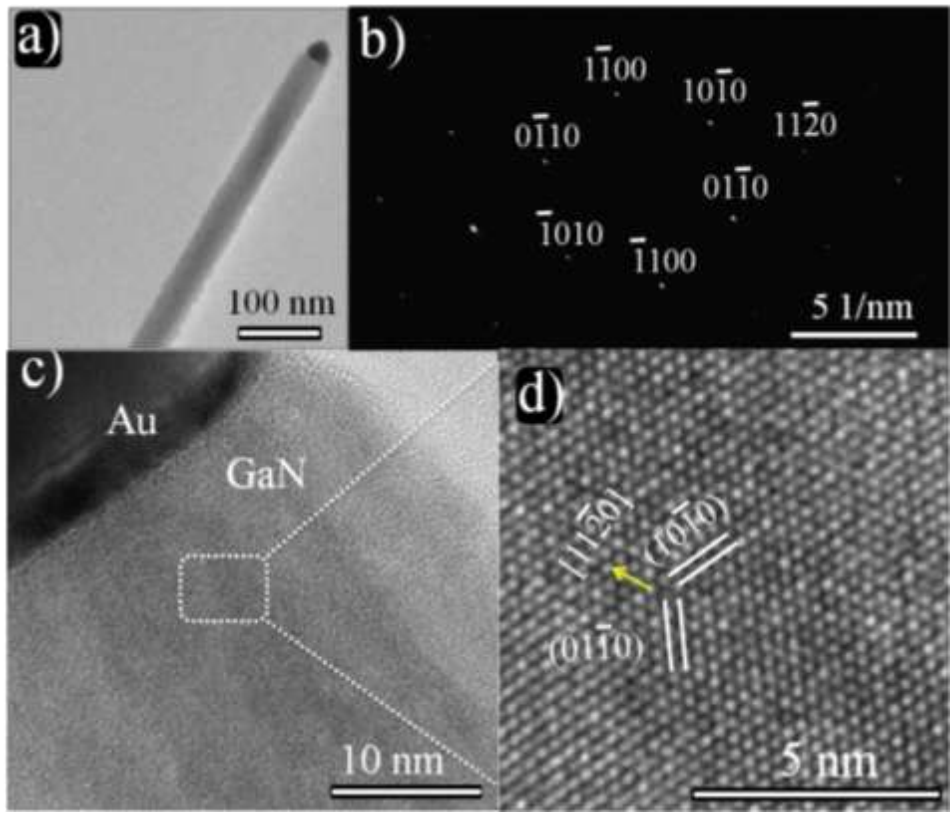

**Fig. 4**. a) Low resolution TEM micrograph of a typical GaN nanowire grown in oxygen reduced condition (sample R3). b) Selected area diffraction pattern of the nanowire, indexed to wurtzite GaN with zone axes along [0001]. c) High resolution TEM image collected near the tip of the nanowire having Au nanoparticle. d) Magnified view of HRTEM, shows the nanowire having nonpolar planes (10$\bar{1}$0) and the growth direction along [11$\bar{2}$0].

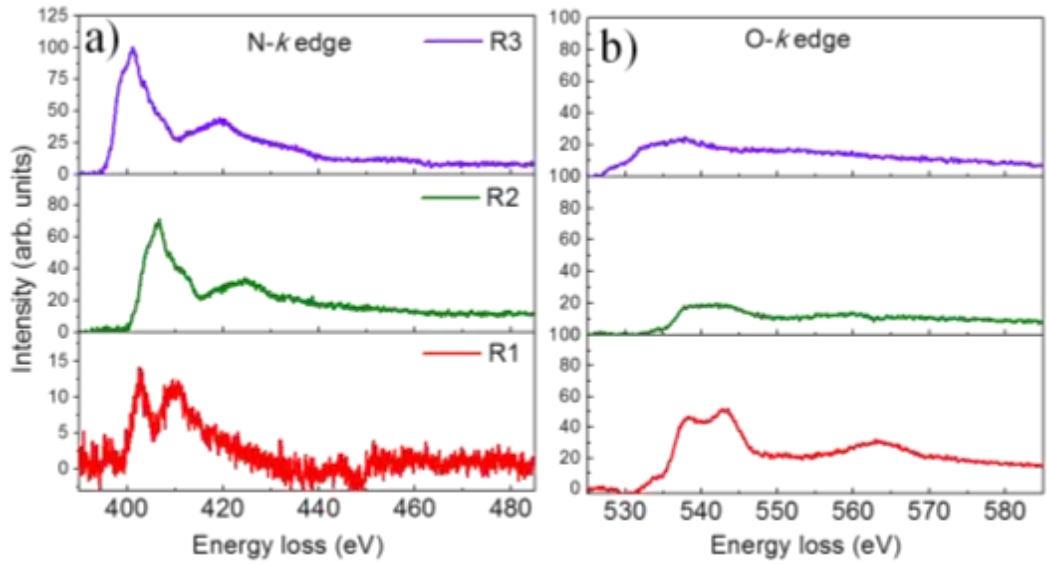

**Fig. 5**. Electron energy loss spectra of GaN nanowires of the samples R1 and R2, R3 grown in O rich and O reduced conditions, respectively. a) Background subtracted N-*k* edge of nanowires. b) Corresponding O-*k* edge of nanowires of the samples R1, R2 and R3.

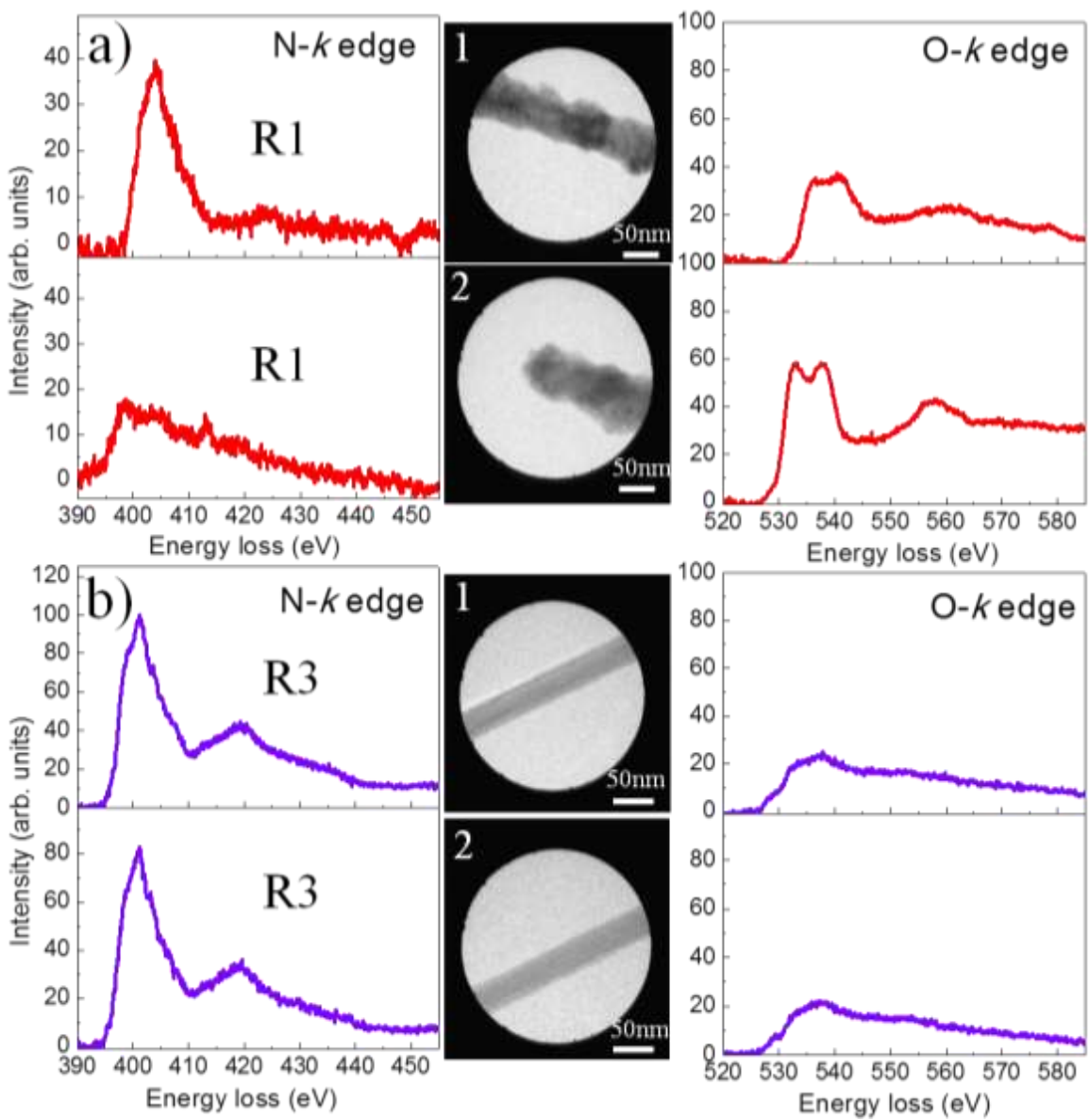

**Fig. 6**. a) EELS spectra of background subtracted *k*-edges of N and O obtained from the position 1 (middle region of the nanowire with thin and rough morphology) and position 2 (tip region of the nanowire with thick and rough morphology) on the nanowire of the sample R1 grown in O rich condition. b) EELS spectra of background subtracted *k*-edges of N and O obtained from the position 1 (middle region of the nanowire) and position 2 (close to one end of the nanowire) on the nanowire of the sample R3 grown in O reduced condition.

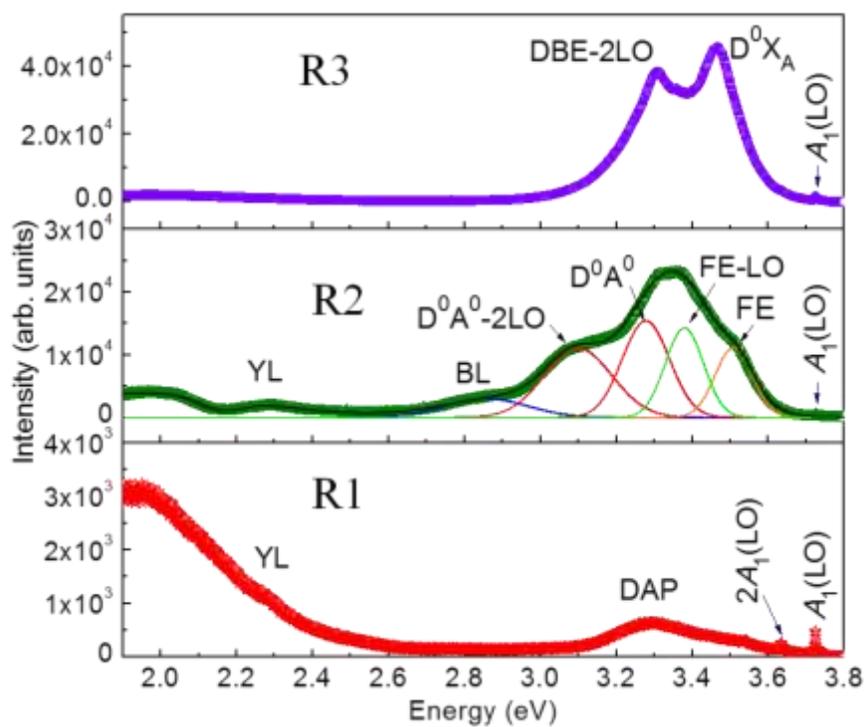

**Fig. 7**. Typical photoluminescence spectra of GaN nanowires of the samples R1, R2 and R3, collected in the energy range 2.0 eV- 3.8 eV, at 80 K.

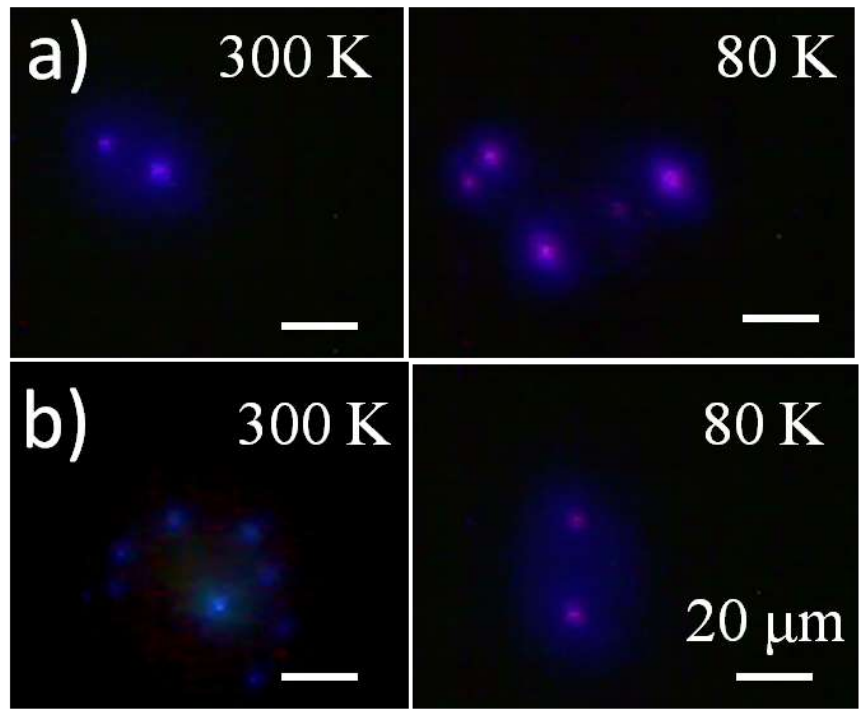

**Fig. 8**. Typical optical images of far field bright violet emission spots at 300 K and 80 K on the nanowires of sample a) R3 and b) R4. Emission spots are coming out of the ends of the nanowires through the available open space while other ends are at the excitation with laser light.

**Supporting Information :**

FESEM images of Au catalyst particles and GaN nanowire with different oxygen impurities

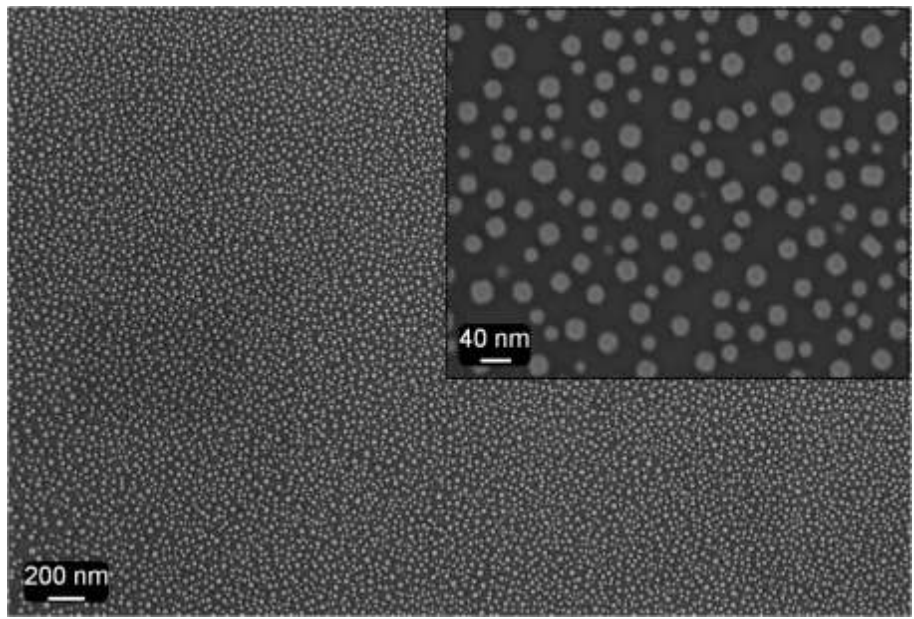

**Fig. S1**. Typical FESEM image of Au nanoparticles on Si (100). Uniform shape and well separated nanoparticles are having the size distribution of 25 – 35 nm. Inset shows magnified view of nanoparticles.

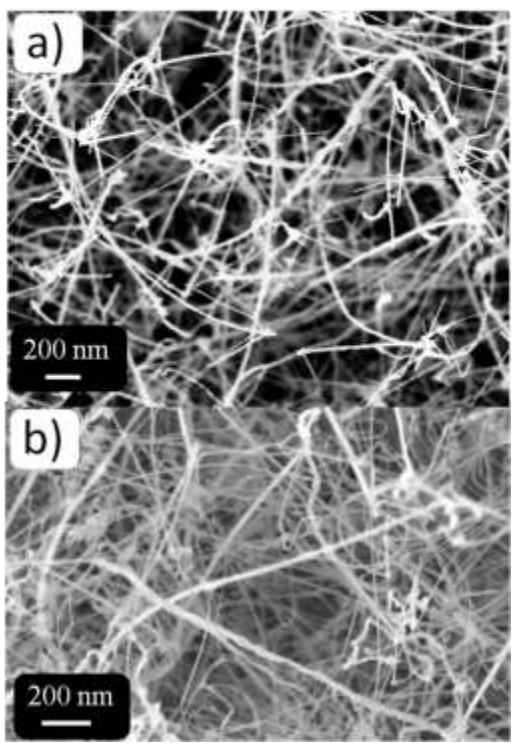

Fig. S2. Typical FESEM images of GaN nanowires with controlled O contamination of a) ~50 ppm and b) $10^4$ ppm showing corrugated morphologies.

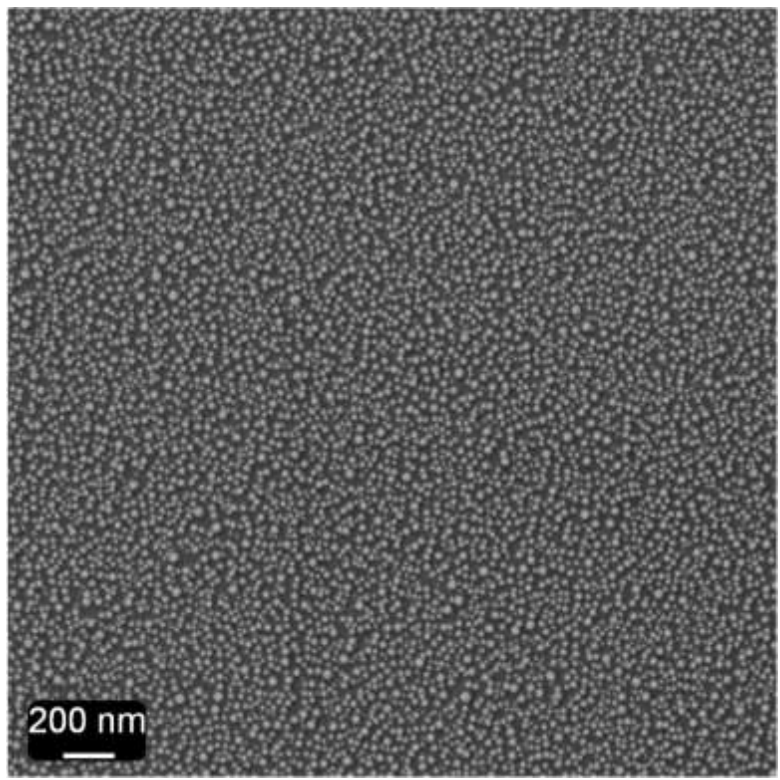

**Fig. S3**. Typical FESEM image of Au nanoparticles grown on Si (100). Nanoparticles are having the size distribution of 35 – 50 nm.